\documentclass{aa}
\usepackage[dvips]{graphics}
\usepackage[english]{babel}
\begin{document}

   \thesaurus{03     
              (11.19.1;  
               11.14.1;  
               11.09.4;  
               09.04.1)}  
   \title{Dust in active nuclei}

   \subtitle{I. Evidence for ``anomalous'' properties}

   \author{R.~Maiolino, \inst{1}
	A.~Marconi, \inst{1}
	 M.~Salvati, \inst{1}
 	 G.~Risaliti, \inst{2}
	 P.~Severgnini, \inst{2}
	E.~Oliva, \inst{1}
	F.~La Franca, \inst{3}
	\and
	 L.~Vanzi \inst{4}
          }

   \offprints{R. Maiolino}

   \institute{Osservatorio Astrofisico di Arcetri
              L.go E. Fermi 5, I-50125, Firenze, Italy\\
              email: maiolino@arcetri.astro.it
	\and
		Dipartimento di Astronomia, Universit\`a di Firenze,
    L.go E. Fermi 5, I--50125, Firenze, Italy
         \and
		Dipartimento di Fisica, Universit\`a degli Studi ``Roma Tre''
   Via della Vasca Navale 84, I--00146, Roma, Italy 
	\and
		European Southern Observatory, Alonso de Cordova 3107,
Santiago, Chile
             }

   \date{Received ; accepted }

\authorrunning{Maiolino et al.}
\titlerunning{Dust in active nuclei}

   \maketitle

   \begin{abstract}
We present observational evidences that dust in the
circumnuclear region of AGNs has different properties than in the
Galactic diffuse interstellar medium. By comparing the reddening
of  optical and infrared broad lines and the X-ray absorbing column
density we find that the $\rm E_{B-V}/N_H$ ratio is
nearly always lower than Galactic by a factor ranging from $\sim$3 up to
$\sim$100. Other observational results indicate that the $\rm A_V/N_H$
ratio is significantly lower than Galactic in various classes of AGNs
including intermediate type 1.8--1.9 Seyferts, hard X--ray selected and radio
selected quasars, broad absorption line QSOs and grism selected QSOs. The lack
of prominent absorption features at 9.7$\mu$m (silicates) and at 2175\AA \
(carbon dip) in the spectra of Seyfert 2s and of reddened Seyfert 1s,
respectively, add further evidence  for dust in the circumnuclear region
of AGNs being different from Galactic.

These observational results indicate that the dust composition
in the circumnuclear region of AGNs could be
dominated by large grains, which make
the extinction curve flatter, featureless and are responsible for the reduction
of the $\rm E_{B-V}/N_H$ and $\rm A_V/N_H$ ratios.

Regardless of the physical origin of these phenomena, the reduced 
dust absorption with respect to what expected from the gaseous column density
should warn about a mismatch between the optical and the X-ray classification
of the active galactic nuclei in terms of their obscuration.

   \end{abstract}

%

\section{Introduction}

The properties of the circumnuclear gas
are a key issue to understand the physics of Active Galactic Nuclei (AGNs).
In particular, gas obscuration has important consequences
on the classification of AGNs (Antonucci 1993), on their infrared
emission (Granato et al. 1997, Pier \& Krolik 1993) and also on the
X-ray background (Setti \& Woltjer 1989). AGNs are, optically-wise, divided
in two main classes: type 1 AGNs, showing broad permitted emission
lines, and type 2 AGNs, which only show narrow emission lines.
The Unified Model assumes that AGNs of both classes host the same kind of
nuclear engine and ascribes their differences solely to orientation effects
with respect to an obscuring gaseous medium, possibly arranged in a
torus-like geometry. For those lines of sights intercepting the obscuring
torus,
both the Broad Line Region (BLR, $< 1$ pc in size) and the nuclear engine
are obscured and only the much more extended
Narrow Line Region (NLR) can be observed. This
model has gained success from a large number of observational tests
(see Antonucci 1993 for a review). In particular, X-ray observations have
supported the unified scenario by discovering large
columns of absorbing gas in type 2 AGNs (eg. Awaki et al.
1991, Turner et al. 1997, Maiolino et al. 1998, Risaliti et al. 1999).
Also, spectroscopic observations in the infrared, where dust absorption is
greatly reduced, detected broad permitted lines in several AGNs which
are classified as type 2 in the optical.

However, the properties of the absorbing medium result to be
more complex with a more quantitative analysis.
Maccacaro et al. (1982) first noted, in a few AGNs,
that the dust reddening affecting the BLR is significantly lower than
expected from the N$_H$ measured in the X-rays, assuming a Galactic
standard extinction curve and dust-to-gas ratio.
Indications for
a low value of $\rm A_V/N_H$ in the obscuring torus
were also found by Granato et al. (1997) by modelling the IR emission
of AGNs; they ascribed the low $\rm A_V/N_H$ ratio to the sublimation of
dust at the inner face of the torus.

The goal of this paper is to observationally verify the
low $\rm E_{B-V}/N_H$ and low $\rm A_V/N_H$
phenomena with a higher confidence (sects. 2 and 3).
We also investigate dust spectral signatures which directly
probe the properties of dust grains in the circumnuclear region of AGNs.
Most of the interpretation of these observational effects is addressed in a
companion paper (Maiolino et al. 2000b, paper II).
Finally,
note that in this paper
we will often distinguish the $\rm E_{B-V}/N_H$ and the
$\rm A_V/N_H$ ratios since, as discussed in paper II, in the
circumnuclear region of AGNs
dust reddening and obscuration are not necessary tied by the Galactic standard
relation ($\rm A_V/E_{B-V}=3.1$).

Throughout this paper we will assume
a cosmology with $\rm H_0=65$ and $\rm q_0=0.5$.


\section{Evidence for a low E$\rm _{B-V}$/N$\rm _H$}

\subsection{Sample selection and reddening determination}
We have defined a sample of AGNs whose X-ray spectrum shows evidence for
cold absorption (hence a measure of the gaseous column density N$_H$
along the line of sight) and whose optical and/or IR spectrum show at
least two {\it broad} lines that are not
completely absorbed by the dust associated to the X-ray absorber.
This sample includes intermediate type (1.8--1.9) Seyferts, type 2 Seyferts
with broad lines in the near-IR, and a few type 1--1.5 Sy characterized
by cold absorption in the X-rays.
We do not consider AGNs whose X-ray
spectrum shows evidence for a warm absorber, since very likely the latter is
not associated to the obscuring torus (Netzer 2000) and also because the
gaseous column inferred for a 
warm absorber is strongly model-dependent (this is discussed
further in Sect.~5).
 We do not consider
cold absorbers with partial covering, since also in this case the absorber is
not associated to the obscuring torus, but to matter very close to the X-ray
source, possibly the BLR clouds (Maiolino 2000). In the cases of dual absorbers
we assume the N$_H$ of the lower column density absorber,
both to be conservative
and because the higher density absorber has generally a low partial covering
(Maiolino 2000).
Tab.1 lists the sources in our sample.

\begin{table*}
\begin{tabular}{lccccccccc}

Name &  E$_{B-V}^a$  & N$_H^b$ & $\rm E_{B-V}/N_H^c$ & z  & Redd. est.$^d$ &
Log L$_X^e$ & \multicolumn{2}{c}{Refs.$^f$} \\
     &               & (10$^{20}$cm$^{-2}$) & (rel. Gal.) &  &
             & (erg s$^{-1}$) & opt/IR & X \\
\hline
NGC4639 & 0.38$\pm$0.25 &  7.3$^{+5.6}_{-5.1}$ &
	3.04$_{-3.02}^{+2.98}$ & 0.00544  & H$\alpha$/H$\beta ^*$ & 40.92
& 1 & 1 \\
NGC5033 & 0.72$\pm$0.24 &  8.7$^{+1.7}_{-1.7}$ & 
	4.84$_{-1.87}^{+1.92}$ & 0.00292  & H$\alpha$/H$\beta ^*$ & 41.10
& 2 & 3 \\                                                                   
M81  &  0.66$\pm$0.25 &  9.4$^{+0.7}_{-0.6}$  & 
	4.11$_{-1.52}^{+1.57}$ & 0.00037  & H$\alpha$/H$\beta ^*$ & 40.52
& 4 & 3 \\                                                                   
SAXJ0045-25 &  0.5$\pm$0.3   &   390$^{+870}_{-260}$ & 
	0.075$_{-0.173}^{+0.067}$ & 0.111    & cont. & 43.06 & 5 & 5
 \\
SAXJ1218+29 &  0.65$\pm$0.2  &  1250$^{+1900}_{-750}$ & 
	0.030$_{-0.047}^{+0.019}$ & 0.176    & cont. & 43.29 & 5 & 5 \\
SAXJ1519+65 &  0.55$\pm$0.2  &  1580$^{+410}_{-320}$ & 
	0.020$_{-0.012}^{+0.007}$ & 0.044    & cont. & 42.74 & 5 & 5 \\
NGC1365 &  1.5$\pm$0.3  & 2000$^{+400}_{-500}$ & 
	0.022$_{-0.005}^{+0.005}$  & 0.005    & Br$\gamma$/Pa$\beta ^*$ & 42.42
& 7 & 3 \\
MCG-5-23-16 & 0.61$\pm$0.59 & 162$^{+23}_{-21}$  &  
 0.220$_{-0.215}^{+0.214}$ & 0.00827     & Br$\gamma$/Pa$\beta$ & 43.22
& 6 & 3 \\                                                   
N5506 &  1.59$\pm$0.59 & 340$^{+26}_{-12}$ & 
	0.27$_{-0.10}^{+0.10}$ & 0.00618     &
Br$\gamma$/Pa$\beta$ & 42.96
& 8,9 & 3 \\                                                   
SAXJ1353+18 &   0.86$\pm$0.15 &  154$^{+37}_{-32}$ & 
	0.33$_{-0.10}^{+0.09}$ & 0.2166   &
Pa$\alpha$/H$\alpha$ & 43.91 & 7,23 & 10 \\
N2992 & 0.58$\pm$0.41 &  90$^{+3}_{-3}$ & 
	0.38$_{-0.27}^{+0.26}$ &  0.0077  & Br$\gamma$/Pa$\beta ^*$ & 43.05
& 11 & 11 \\
AXJ0341-44 &  $<$1.59       & 1000$^{+400}_{-400}$ & 
	$<$0.09  & 0.672   & H$\beta$/H$\gamma ^*$ & 44.21
& 12 & 13 \\
Mkn6  &   0.72$\pm$0.13 &  333$^{+29}_{-16}$ & 
	0.13$_{0.02}^{+0.02}$ & 0.01847  & H$\alpha$/H$\beta ^*$ & 43.17
& 14 & 14 \\                                                   
3C445  & 0.88$\pm$0.21 &  580$^{+320}_{-180}$ & 
	0.088$_{-0.053}^{+0.034}$ & 0.057   & H$\alpha$/H$\beta ^*$ & 43.79
& 15 & 16 \\                                                   
PG2251+11 & 0.34$\pm$0.13 &   60$^{+30}_{-22}$ & 
	0.33$_{-0.21}^{+0.17}$ & 0.3255   & H$\alpha$/H$\beta ^*$ & 44.67
& 17 & 16 \\                                                   
IRAS13197-1627 & 0.47$\pm$0.15 & 3000$^{+1100}_{-1000}$ & 
	0.0092$_{-0.0044}^{+0.0042}$ & 0.01654  &
H$\alpha$/H$\beta ^*$  & 42.66 
& 18 & 19 \\ 
IRAS+0518 &  0.79$\pm$0.56 & 840$^{+80}_{-65}$  & 
	0.055$_{-0.039}^{+0.039}$ &  0.042    & Pa$\alpha$/Pa$\beta ^*$ & 43.44
& 19 & 19 \\
NGC526a & 1.0$\pm$0.25 &  150$^{+14}_{-14}$ & 
	0.39$_{-0.10}^{+0.10}$ & 0.01922  & Pa$\beta$/H$\alpha$  & 43.55
& 7,22 & 3 \\                                                   
Mk231 & 0.34$\pm$0.10 &  370$^{+260}_{-150}$ & 
	0.054$_{-0.041}^{+0.027}$ & 0.042 &  Pa$\alpha$/H$\alpha ^*$ & 42.66
& 21 & 20  \\ 
\hline
\end{tabular}

\caption{Sample of AGNs whose X-ray spectrum shows evidence for cold
absorption and whose optical and/or IR spectrum shows at least two
broad lines. Notes: $^a$ reddening estimated assuming a standard Galactic
extinction curve; $^b$ gaseous column density inferred from the hard X-rays
in units of 10$^{20}$ cm$^{-2}$; $^c$ E$_{B-V}$/N$_H$ ratio relative to the
standard Galactic value of 1.7$\times$10$^{-22}$cm$^{-2}$; $^d$ lines ratio
used to estimate the reddening (``cont.'' indicates objects for which the
reddening was estimated by fitting also the continuum), stars mark line pairs
which were observed (nearly) simultaneously; $^e$ 2--10 keV {\it intrinsic}
luminosity (i.e. corrected for gaseous absorption);
$^f$ references for the optical,
infrared and X-ray data: 1) Ho et al. (1999), 2) Koratkar et al. (1995),
3) Bassani et al. (1999), 4) Filippenko \& Sargent (1988), 5) Maiolino et al.
(2000a), 6) Veilleux et al. (1997), 7) Marconi et al. (in prep.),
8) Rix et al. (1990), 9) Maiolino et al. (1996), 10) Vignali et al. (2000),
11) Gilli et al. (2000), 12) Halpern et al. (1999), 13) Boyle et al. (1998),
14) Feldmeier et al. (1999), 15) Crenshaw et al. (1988), 16) Sambruna et al.
(1999), 17) Stirpe et al. (1990), 18) Aguero et al. (1994),
19) Severgnini et al. (in prep.), 20) Turner et al. (1999),
21) Lacy et al. (1982), 22) Winkler et al. 1992, 23) Fiore et al. 1999.}

\end{table*}

\begin{figure}[h]
\resizebox{\hsize}{!}{\includegraphics{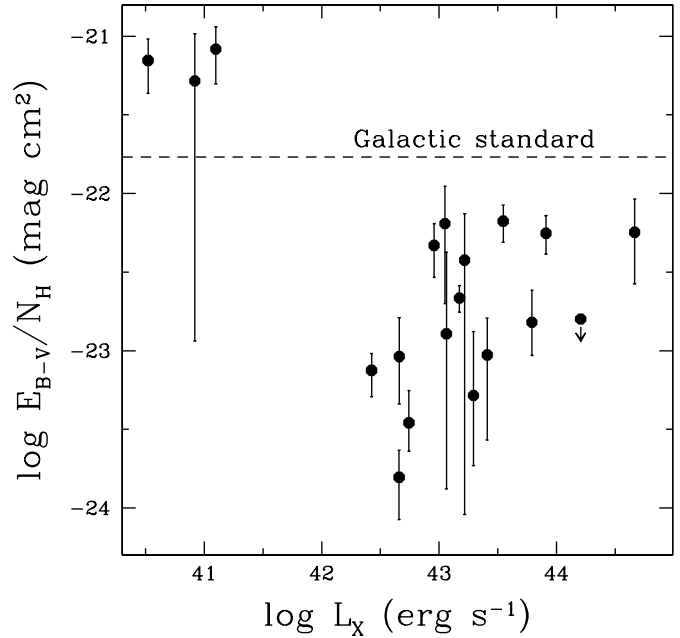}}
\caption{$\rm E_{B-V}/N_H$ ratio as a function of the intrinsic 2--10 keV
luminosity for the objects in our sample. The dust reddening
$\rm E_{B-V}$ is estimated assuming
a Galactic standard extinction curve as discussed in the text.
The gaseous column density $\rm N_H$ is derived from the photoelectric
cutoff in the hard X-rays.}
\end{figure}

The gaseous N$_H$ along the line of sight
is derived directly by the photoelectric
cutoff in the X-ray spectrum, provided that the signal-to-noise is high
enough. Tab.~1 lists the N$_H$ derived for the sources in our sample
along with the {\it intrinsic} 2--10 keV luminosity (i.e. absorption corrected).
As mentioned above we excluded objects whose X-ray spectrum shows
evidence for warm absorption as based on the presence of the absorption edges
of OVII and OVIII at 0.74 and 0.87 keV respectively.
For most of the objects
in Tab.~1 the X-ray absorption is ``cold'', in the sense that the chi-squared
of the  spectral
fit is significantly better with the latter model than
with a warm absorption model (this is discussed in some of the references
reported in Tab.~1). For some objects
the signal-to-noise of the X-ray spectrum is not high enough to
discard warm absorption; this is certainly the case for
SAX0045-25, SAX1218+29, SAX1519+65, AXJ0341-44, and
Mkn231. Yet, the spectral shape in most of the latter cases is such that a fit
with a warm absorber would require an unrealistically high colum of warm gas
($\rm \sim 10^{24}cm^{-2}$, possibly with the exception of Mkn231) thus
favoring the cold absorption model. Finally, even in those cases
for which cold absorption provides a better fit with respect to a warm
absorber this generally does not exclude that both components are present.
Also, ionized gas along the line of sight might not be sampled by
the OVII and OVIII absorption edges if the ionization stage is too low or too
high (Kraemer et al. 1999, 2000, Brandt et al. 1996, Reynolds \&
Fabian 1995). However, the presence of ionized gas along the line of sight
which is not detected in the X-ray spectra would make the total column
of gas (neutral+ionized) higher than inferred assuming a single cold,
component and, as we shall see, the problem of the reduced
$\rm E _{B-V}/N_H$ would be even worst.

Ratios between broad components of the hydrogen lines compared
to the intrinsic values give the amount of dust reddening affecting
the BLR. However, radiative transport and collisional excitation effects
in the extreme conditions of the BLR clouds ($\rm n\sim 10^9 cm^{-3}$)
can affect the standard hydrogen line ratios expected in the case B
recombination. For instance, BLR models expect the H$\alpha$/H$\beta$ Balmer
decrement to range
from the ``standard'' ratio of 3.1 up to a factor of 3 higher (Rees et al.
1989, Netzer et al. 1985,  Mushotzky \& Ferland 1984), this
is mostly due to the large optical thickness of H$\beta$ whose
de-excitation
 transition has a high probability of being split into Pa$\alpha$ and
H$\alpha$.
Nonetheless, the H$\alpha$/H$\beta$ ratio observed in Sy1s and QSOs is often
consistent with the standard case B value. Should the
intrinsic ratio be higher, the observed H$\alpha$/H$\beta$ compared to the
case B
value provides at least an upper limit to the reddening.
Ratios between infrared broad hydrogen lines
(Pa$\beta$, Pa$\alpha$ and
Br$\gamma$) provide a more reliable measure
of the reddening, since they are much less
affected by the radiative transport effects discussed above.
In some cases the broad lines were not measured
simultaneously and therefore variability might have affected the real ratios.

As a consequence of what discussed above we adopted the following criteria to
select the broad line pairs to be used for the reddening determination.
When  more than two
broad lines were available, to avoid problems related to the variability
 we chose those lines which
were observed (nearly) simultaneously.
The cases for which the broad lines were measured simultaneously
are marked with a ``*'' on column 6
of Tab.~1 (these cases are the majority). When more than two simultaneous broad
lines were available we used the 
ratios involving only near-IR lines rather than
H$\alpha$/H$\beta$. Finally,
Pa$\beta$/H$\alpha$,
Pa$\alpha$/H$\alpha$ and (in one case) H$\beta$/H$\delta$ were
used only in a few cases.
The line ratio adopted to estimate the reddening for each object is given on
column 6 of Tab.~1.

To determine the reddening E$_{B-V}$ from the broad line ratios an extinction
curve must be assumed. We have assumed
the ``standard'' Galactic extinction curve (Savage \& Mathis 1979).
To derive the extinction we also assumed a foreground screen,
which is a reasonable assumption given that the BLR
is within the sublimation radius (and therefore mostly dust-free) and
that the torus is probably more extended than the BLR (paper II).
Column 2 of Tab.~1 lists the dust reddening
toward the BLR derived for each object.

We also include in our study three AGNs (namely
SAXJ0045-25, SAXJ1218+29, SAXJ1519+65) whose reddening was
not inferred from the ratios of broad hydrogen lines but by a detailed fit
to the continuum and of the EW of the H lines. More specifically,
the optical spectrum (including shape, stellar features and broad H lines),
along with near-IR to U-band spectroscopic measurements, were fitted 
with stellar population synthesis templates of various ages combined with
an AGN template reddened by various degrees of extinction. The three 
free parameters required to fit the data were the relative contribution
of the stellar and of the AGN component, the age of the stellar population and
the reddening of the AGNs. For a more detailed description of the method
see Maiolino et al. (2000a) and Vignali et al. (2000). Although the large
number of observational constraints allow a relatively good determination of
the three free parameters, the determination of the reddening
is not as accurate as for the broad lines ratio method. However, for one object
(SAXJ1353+18) both methods could be used and gave consistent results
(Maiolino et al. 2000a, Vignali et al. 2000).
The interesting property of these objects is that they were selected
(and actually discovered) in the hard X-rays (Fiore et al. 1999)
and, therefore, are not
affected by some of the biases that might affect the other objects and which
will be discussed later.

\subsection{Evidence for a low $\rm E_{B-V}/N_H$}

On column 4 of Tab.1 we report the $\rm E_{B-V}/N_H$ ratio relative to the
Galactic standard value of $\rm 1.7\times 10^{-22}
mag~cm^2$ (Bohlin et al. 1978).
Except for a few cases, $\rm E_{B-V}/N_H$
is significantly lower than the Galactic
standard value, by a factor ranging from a few to $\sim$100.
This is graphically shown in Fig.1 where the $\rm E_{B-V}/N_H$
ratio relative to Galactic is plotted as a function of the intrinsic X-ray
luminosity. Fig.1 does not really show a correlation between the two
quantities, but rather a bimodal behavior: 
AGNs with luminosities higher
than $\rm 10^{42} erg~s^{-1}$ are systematically
characterized by $\rm E_{B-V}/N_H$ lower
than Galactic, though they show a large spread, while those few
Low Luminosity AGNs
(LLAGNs) in our sample,
with $\rm L_X < 10^{42} erg~s^{-1}$, are characterized by $\rm E_{B-V}/N_H$
consistent with, or even higher, than Galactic. The markedly different
 behavior of LLAGNs
with respect to the other AGNs in the sample, might reflect the fact that
the physics of these objects is intrinsically different from the ``classical''
AGNs, as suggested by various authors (eg. Ho 1999).
In the following we will focus on the other AGNs in our
sample, which have luminosities more typical of classical Seyfert galaxies or
of QSOs.

Admittely,
our sample is not very large and is not representative of the
population of obscured AGNs, given that we had to select our objects in a
relatively narrow range of absorptions. In particular, the absorption must
be low enough to  
enable us to detect the broad lines. We cannot exclude the existence of
AGNs (in the normal-high luminosity range $\rm > 10^{42}erg~s^{-1}$) with
an $\rm E_{B-V}/N_H$ consistent with Galactic. For instance, objects with
$\rm N_H$ higher than a few times $\rm 10^{22}cm^{-2}$ and {\it Galactic}
$\rm E_{B-V}/N_H$ would be characterized by an extinction $A_V$ so high to
make broad lines undetectable, possibly even in the IR, and therefore would
be excluded from our sample. Nonetheless, we would like to stress the
following results: 1) the existence of a population of AGNs characterized
by a value of $\rm E_{B-V}/N_H$ significantly lower than Galactic
is proven beyond any doubt;
2) with the exception of the three LLAGNs discussed above, we could not
find AGNs whose $\rm E_{B-V}/N_H$ is consistent with the Galactic
standard value.\footnote{Note that even with the bias discussed above we would
have expected to find some object with Galactic $\rm E_{B-V}/N_H$, at least in
those cases with $\rm N_H \ge 10^{22}cm^{-2}$ and observed in the near-IR.}
Determining if this is a property common to all AGNs or determining what
fraction of them is characterized by this feature cannot be done with the
current sample alone.

\section{Evidences for a low A$_V$/N$_H$}

The evidence for a low $\rm E_{B-V}/N_H$ discussed above might be
translated into evidence for a low $\rm A_V/N_H$ if a standard
conversion factor $\rm A_V/E_{B-V} = 3.1$ applies to these objects.
Indeed, such a conversion factor might not apply
and, therefore, a reduced reddening does not necessary
imply a reduced absorption. However, there are other observational
evidences supporting the idea that also the A$_V$/N$_H$ ratio AGNs
is generally significantly lower than the Galactic standard value.

\subsection{The intermediate 1.8--1.9 type Seyferts}

The intermediate type Seyferts discussed in the former section are only
a small fraction of those available in various Seyfert samples; in
particular the discussion in Sect.2 is limited to
the sources for which enough information is available to derive both
reddening and gaseous column density. In some of these objects
 the weakness of the
broad H$\beta$ or H$\alpha$ lines is due to intrinsic properties of the BLR
(Goodrich 1995), but more often
the weakness of the broad lines lines is ascribed to dust absorption (Maiolino
\& Rieke 1995). In the latter case
the extinction must be about $\rm A_V \approx 3$, given that usually in these
objects the
faint broad lines have a flux similar to the narrow components, while
in type 1 objects broad lines are about 10 times stronger than the
narrow lines. On the other hand, most of the intermediate type Seyferts
are characterized by an absorbing column density between $10^{22} cm^{-2}$
and $10^{23} cm^{-2}$ and, in some cases,
 even higher than $10^{23} cm^{-2}$ (Risaliti et al. 1999).
With a Galactic conversion factor such column densities would imply
an extinction of $\rm A_V \approx 30$, which would obviously make any
broad line undetectable (unless reflected, paper II). This suggests
that, at least in this class of objects, the A$_V$/N$_H$ ratio must
be about a factor of 10 lower than Galactic.

\begin{figure}[h]
\resizebox{\hsize}{!}{\includegraphics{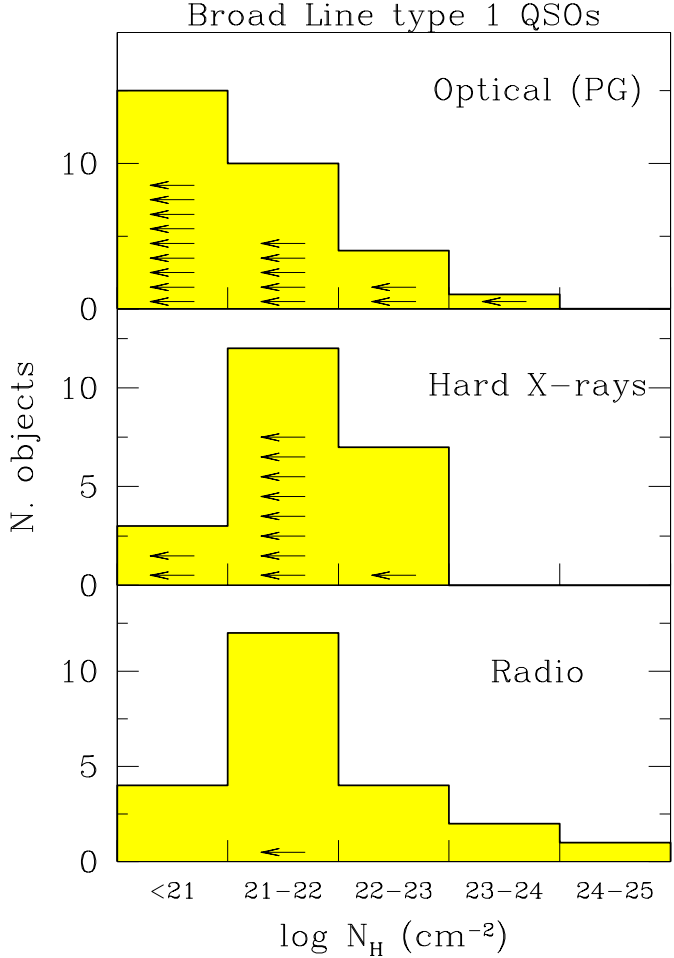}}
\caption{Distribution of gaseous absorbing column density, as inferred from
hard X-ray observations, for various samples of type 1, broad line AGNs, more
specifically:
a subsample of the PG QSOs (top), a
sample of hard X-ray selected QSO (middle) and a sample of radio selected
AGNs (bottom).
}
\end{figure}

\subsection{QSOs with hard X-ray absorption}

Local Seyfert 1 galaxies are generally characterized by an X-ray spectrum
with low or no cold absorption, in agreement with the extinction
inferred by their optical spectra. However, at higher luminosities, there
is evidence for a population of QSOs whose hard X-ray
spectrum is characterized by significant
cold absorption along the line of sight although their optical spectrum
has prominent broad lines and, usually, also a blue continuum typical of
unabsorbed AGNs.

In particular, although most of the
optically selected PG QSOs do not show evidence for X-ray
gaseous absorption in excess of the Galactic value, a few of them are
affected by an excess absorption with $\rm N_H > 10^{21} cm^{-2}$.
This is shown in the top panel of Fig.~2, where the distribution of
cold absorbing columns, as inferred from the hard X-rays, is
reported for the PG QSOs.

Obviously, even a small amount of dust associated with the gaseous column
along the line of sight can redden the QSO continuum enough to be missed
by U--B color selection criterion of the PG survey. Indeed,
surveys of type 1 QSOs selected at wavelength less sensitive to
dust obscuration and reddening
have found a larger fraction of objects which show
evidence for significant gaseous absorption. In particular, the lower
two panels of Fig.~2 show the distribution of N$_H$ for the type 1 QSOs in
the ASCA survey presented by Akiyama et al. (2000)
 and in the radio-selected type 1
AGNs presented by Sambruna et al. (1999). These distributions clearly
show a larger fraction of type 1 AGNs with N$_H$ even in excess of
$\rm 10^{22} cm^{-2}$. On the other hand, the prominent broad lines and,
often, the blue continuum (although not as ``blue'' as for the PG QSOs)
suggest that the optical absorption and reddening cannot be large.
In particular, Akiyama et al. (2000) show that, although significant X-ray
absorption is observed in their ASCA selected objects, the ratio between
optical and hard X-ray emission is not different from soft X-ray
selected QSOs (i.e. not affected by significant
absorption). This indicates
that optical absorption must be lower than about 1 mag\footnote{The hard X-ray
flux is not significantly affected by absorption for N$_H$ values which
are typical of these objects ($\rm N_H < 10^{23} cm^{-2}$, Fig.~2).}
and, together with the N$_H$ inferred from the X-rays,
indicates an $\rm A_V/N_H$ about a factor of ten lower than Galactic.

The HELLAS BeppoSAX survey (Fiore et al. 1999, Comastri et al. 2000) has also
found some type 1 blue QSOs with flat X-ray spectra suggesting large columns of
gas along our line of sight. Although statistically less relevant than the
surveys mentioned above, we could study some of these blue X-ray--absorbed
QSO more in
detail (eg. Maiolino et al. 2000): we found an $\rm E_{B-V}$
 which is a factor of
$\sim$30 lower than expected from the X-ray absorbing
 N$_H$ and a Galactic $\rm E_{B-V}/N_H$.

Within this context it is puzzling that, at variance with the ASCA and
BeppoSAX results,
Chandra has not found a large number of objects of this class
so far (Fiore et al. 2000, Brandt et al. 2000b, Mushotzky et al. 2000):
only a few blue type 1 QSOs with significant N$_H$ were discovered by Chandra's
surveys. Possibly this is due to the sensitivity of Chandra which peaks below 2
keV and, therefore, probably biases the results  in favor of 
soft, little absorbed X-ray sources. XMM, whose sensitivity is much more
uniform up to $\sim 7 keV$, should tackle this issue.

\subsection{Broad Absorption Line QSOs}

Probably the most extreme case of type 1 AGNs with low $\rm A_V/N_H$
are the Broad Absorption Line (BAL) QSOs. Brandt et al. (2000a) and 
Gallagher et al. (1999)
found that these objects are extremely weak both in soft X-rays and
hard X-rays with respect to the optical.
They suggest that this property is not due their intrisic SED,
but to a large column of absorbing gas along the line of sight, probably related
to the same medium seen in UV resonant absorption lines.
The X-ray emission, heavily suppressed even in the
2--10 keV band, suggests that the gaseous absorbing column must be relatively
high:
$\sim 10^{23} cm^{-2}$ or higher. Yet, the optical to UV spectrum is rather
typical of normal, unabsorbed QSOs (except for the presence of the broad
resonant absorption lines), implying little or no dust absorption.
Combined with the large absorbing column inferred from the X-rays, this
implies that these objects are characterized by an $\rm A_V/N_H$ ratio
which is nearly two orders of magnitude lower than Galactic.

\subsection{Grism selected QSOs}

As mentioned above, even a small amount of dust might redden the QSO
spectra enough to exclude them from surveys based on color selection
criteria such as the Palomar Green survey.
On the other hand, grism QSO surveys are mostly
based on the detection of broad lines whose flux might be little
affected by extinction in the case of a low
 $\rm A_V/N_H$, even if a substantial amount of gas were present
along the line of sight. Therefore, grism-selected QSOs are among
the best suited objects
to search for effects of a low $\rm A_V/N_H$ ratio. Indeed,
by comparing the optical and X-ray emission of grism selected QSOs,
 Risaliti et al. (2000) find evidence
for a significant population of QSO characterized by a low $\rm A_V/N_H$ ratio.
The reader is addressed to that paper for an exhaustive discussion.

\section{Evidences for peculiar properties of dust grains}

In the former two sections we have discussed observational evidences 
on a reduced dust reddening and absorption towards active nuclei
with respect to what would be
 expected from the gaseous N$_H$, assuming a Galactic
gas--to--dust ratio and extinction curve.
 This discrepancy can be ascribed to various effects,
as discussed in detail in paper II. One
possibility is that the properties of dust grains in the circumnuclear region of
AGNs are different from the diffuse ISM of our Galaxy. In the following we
present two observational evidences supporting this scenario.

\subsection{The 9.7$\mu$m silicate feature}

The silicate feature at 9.7$\mu$m observed in the mid-IR spectra of
many Galactic sources is commonly ascribed to silicate grains with
sizes smaller than $\sim 3 \mu$m.
Ground-based mid-IR studies of type 2 Seyferts claimed
the detection of a deep silicate absorption feature
 in the spectra of many
Sy2s, which was regarded as an evidence supporting the unified model.
However, recent ISO spectra have shown that many of such detections were
probably spoilt by the narrow bandwidth of the data, limited
by the atmospheric transmission. Indeed, the presence of strong PAH
features on both sides of the silicate dip prevented a reliable determination
of the continuum in ground-based observations. The average ISO spectrum
of a sample of Sy2s obtained by Clavel et al. (2000) does not show evidence
for any silicate dip, thus questioning the case for a significant
absorption feature claimed in former studies. In particular the
very conservative upper limits on the EW of the silicate feature given by
Clavel and collaborators for the average spectrum of Sy2s (EW$<$0.32$\mu$m)
implies $\tau _{9.7} < 0.6$.

On the other hand the mid-IR continuum produced
by the active nucleus is
certainly suppressed in Sy2s with respect to Sy1s. This is apparent, as
discussed by Clavel et al. (2000),
in the much larger equivalent width of the PAH
features of Sy2s with respect to Sy1s. In particular, Clavel et al. estimate
an average dust extinction in the mid-IR band and, more specifically,
at 7.7$\mu$m of about 1.83 mag with a dispersion
of $\pm$0.74 mag. For a Galactic dust composition this would give
$\tau _{9.7} = 6.2\pm 2.5$, i.e. one order of magnitude
higher than the upper limit of 0.6 derived from the constraints on the silicate
absorption feature, as discussed above.

Note that the absorptions derived above assume a uniform
foreground screen model. In the reality the emitting region is extended ($\sim$
10 pc) and mixed with the absorber, though most of the emission comes from the
hot, inner region. As a consequence, the optical depths derived above actually
gives an ``equivalent'' optical depth for screen absorption.
Nonetheless, the relation between $\tau _{9.7}$ and $\tau _{7.7}$
does not change much with the geometry of the absorber,
since we are comparing two different forms of absorption of the
same continuum at nearly the same wavelength.
In particular, the relation remains unchanged in the case of partial
covering of the absorber (eg. patchy extinction) since the optical depth
at 7.7$\mu$m and at 9.7$\mu$m are measured by means of the equivalent width of
two features at nearly the same wavelength. In the case that the emitting
region is mixed with the absorber the expected depth of the silicate feature
is lower. If the emitting
region is completely mixed with the absorber (which is an extreme case
since the colder absorbing gas must be located in the outer
parts with respect to the mid-IR nuclear emitting region)
the ``equivalent'' optical depth in the silicate feature inferred
from the (equivalent) $\tau _{7.7}$ should be $\tau _{9.7} = 3.1\pm 0.68$
which is still much higher than the upper limit given by the non-detection
of the silicate feature in abosorption.
Therefore, althogh the meaning of the optical depth derived at 7.7$\mu$m and
at 9.7$\mu$m depends on the geometry of the circumnuclear emitting/absorbing
matter, the discrepancy between these two quantities remains significant and is
not to ascribe to geometrical effects.
The only concern to this regard, is that
the inner hot region, which is absorbed along our line of sight by the outer
colder regions, might be characterized by the silicate feature in emission, as
it is observed in Sy1s (Clavel et al. 2000). Since the upper limit on $\tau
_{9.7}$ was based on the upper limit on the equivalent width of the 9.7$\mu$m
absorption feature assuming an intrinsic featureless continuum, the possible
presence of the emission feature implies a higher upper limit on $\tau
_{9.7}$. However, even in Sy1s the 9.7$\mu$m silicate emission is relatively
faint (EW$\approx 0.25 \mu$m) and, in the worst case, accounting for this
emission feature would only 
increase the upper limit on $\tau _{9.7}$ by 0.35,
i.e. $\tau _{9.7} < 0.95$, which is still much lower than what was derived
from the featureless absorption at 7.7$\mu$m ($\tau _{9.7} = 4.8$).

This result is a very convincing
indication that although dust must be absorbing the mid-IR nuclear radiation,
such dust must have properties different from the Galactic
diffuse interstellar medium.
As discussed in paper II the most likely explanation is that dust in the
circumnuclear region is
predominantly composed of grains with size larger than
3$\mu$m, which do not contribute to the feature at 9.7$\mu$m.

\begin{figure}[h]
\resizebox{\hsize}{!}{\includegraphics{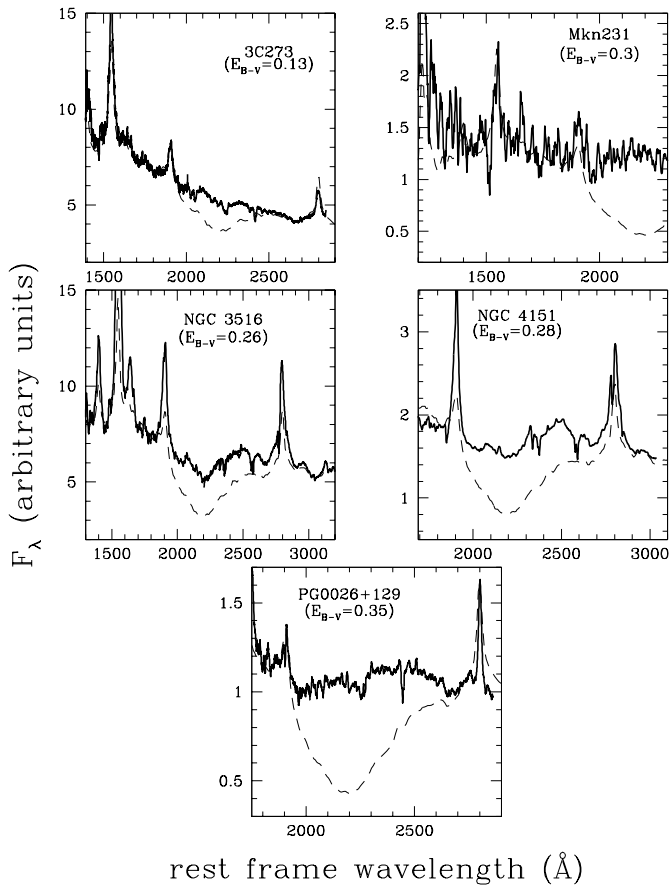}}
\caption{UV spectra of four type 1 AGN whose broad lines ratios
and continuum suggest dust absorption along the line of sight.
The thin dashed line is the average spectrum of type 1 AGNs
(mostly QSOs) reddened with a standard Galactic extinction curve
with an $E_{B-V}$ consistent with that inferred from the broad lines
ratio and adapted (within the uncertainties of the $\rm E_{B-V}$
measured with the broad lines) to match the shape of the continuum in those
regions not affected by the carbon dip around 2175\AA. Note that the
Galactic extinction curve always predicts a significant absorption
by this feature which is not observed.
}
\end{figure}

\subsection{The 2175\AA \ carbon dip}

The absorption feature
 at 2175\AA \ observed in the diffuse interstellar medium was
commonly ascribed to small graphite grains with radii $\approx 100-200$\AA .
More recent studies ascribe most of the profile of
this absorption feature to even smaller dust
particles (PAHs, Weingartner \& Draine 2000).
Observing this feature in absorbed AGNs is really difficult since even a small
amount of dust generally suppresses nearly completely the UV emission. On the
other hand unabsorbed AGNs are generally free of any dust absorption.
Yet there are a few type 1 Seyferts and
QSOs whose optical and near-IR broad line ratios
suggest the presence of some dust reddening along the line of sight (Lacy et
al. 1982, McAlary et al. 1986, Puetter et al. 1981).
Some of these objects are still relatively bright in the UV and have been
observed with the HST spectrometers (FOS and STIS); these are therefore among
the best suited objects to look for the carbon dip feature at 2175\AA .  
In Fig.3 we show the UV spectra of five of these slightly reddened type 1
AGNs for which we could retrive HST archival spectra. Since we are
mostly interested in the continuum shape and in the broad carbon absorption
feature the spectra
were smoothed to a resolution of about 1000 km/s. The thin
dashed line shows the template of type 1 AGNs obtained by Francis
et al. (1991) reddened with  the standard Galactic extinction curve by an
amount consistent with an $\rm E_{B-V}$ consistent both
 with the reddening of  the broad lines and with the
shape of the UV continuum
outside the carbon dip.
The most important result is that the Galactic
extinction curve sistematically predicts a deep feature around 2175\AA \ which
is undetected or much weaker in the observed
spectra \footnote{Note that some of the objects show a small FeII bump
at 2500\AA \  which is stronger than in the template; however, such an emission
feature cannot account for the missing dip expected at 2175\AA .}

As discussed in paper II, this observational evidence supports the
idea that small grains are depleted in the dusty medium responsible for the
reddening.


\section{Evidences for $\rm E_{B-V}/N_H$ higher than Galactic?}

Finally, we shall discuss whether there is any evidence in other previous
studies for an $\rm E_{B-V}/N_H$ higher than Galactic in any AGNs,
besides the three LLAGNs discussed  in Sect.2.

Veilleux et al. (1997) found a few
AGNs whose $\rm A_V/N_H$ is higher than Galactic (some of which are also
included in our sample). For most of these objects a high lower limit to $\rm
A_V$ was obtained by comparing a broad Pa$\beta$ with an upper limit on the
broad component of H$\alpha$. Paradoxically, as acknowledged by the same
authors, for the same objects the comparison between broad Pa$\beta$ and broad
Br$\gamma$ gives little or no reddening. We believe that this result
reflects problems with estimating the upper limits for the broad
component of H$\alpha$ which, most probably, was underestimated.
The only object in their paper for which $\rm A_V/N_H$ seems
larger than Galactic and for which they do not use upper limits on the broad
line components is NGC2992. However,
 the much higher quality optical and near-IR
(simultaneous) spectra obtained by Gilli et al. (2000) clearly indicate an
absorption significantly lower than inferred by Veilleux et al. (1997).

A significant fraction of Sy1s shows evidence for highly ionized gas along our
line of sight, identified  through the presence of absorption edges of OVII and
OVIII in the soft X-rays,
referred to as ``warm absorber'' (Reynolds \& Fabian 1995, George et al. 1998). In
some cases the X-ray spectra also show evidence for neutral gas in addition to
the highly ionized gas. For some of these Sy1s with warm absorbers there is
also evidence for dust reddening (Komossa \& Fink 1997a,b, Leighly et al. 1997,
Komossa \& Bade 1998, Reynolds \& Fabian 1995). The comparison
between the $\rm E_{B-V}$ and the column of neutral gas
suggesting an $\rm E_{B-V}/N_H$ higher than
Galactic. However, we note that there is no reason for the highly ionized gas
to be dust-free if it is located at a distance larger than the sublimation
radius. Indeed, the columns inferred for the highly ionized component are in
agreement (or higher) with the dust reddening for a Galactic dust--to--gas
ratio. Secondly, as pointed out by various authors, the ionized gas probably
has multiple components at different ionization stages, which are not sampled
by the OVII and OVIII edges, but which are characterized by columns high enough
to account for the observed reddening (Kraemer et al. 1999,
2000, Brandt et al. 1996,
Reynolds \& Fabian 1995).

\section{Conclusions}

We have reported various observational evidences indicating that the dust
reddening and absorption of the nuclear region of AGNs
is generally much lower than the values expected from the gaseous column
density measured in the X-rays, if a standard Galactic dust--to--gas ratio
and extinction curve are assumed.

Quantitatively, the most convincing argument supporting a low
$\rm E_{B-V}/N_H$ ratio is the comparison between the reddening inferred from
the optical and infrared broad line ratios and the N$_H$ derived from the
hard X-rays for a sample of 20 objects.
With the
exception of three low luminosity AGNs ($\rm L_X < 10^{42} erg~s^{-1}$),
whose physics might differ from Seyfert-- and QSO--like luminosity systems,
all the objects appear characterized by an $\rm E_{B-V}/N_H$ ratio
sistematically lower than the Galactic standard value by a factor ranging from
$\sim$3 up to $\sim$100.

Also, we have presented
additional evidences suggesting a reduced $\rm A_V/N_H$ ratio in AGNs.
The presence of substantial gaseous absorbing
column ($\rm N_H \sim$ a few times $10^{22}$) in intermediate
type 1.8--1.9 Seyferts
contrasts with their optical appearance, but can be
reconciled by a low $\rm A_V/N_H$ ratio. A more extreme case of this
effect is observed in a number of type 1 AGNs (mostly QSOs) whose X-ray
spectrum shows evidence for substantial gaseous absorption despite their
optical unabsorbed appearance. Two classes of type 1 AGNs which seems to be
nearly systematically affected by this phenomenon
(i.e. significant X-ray absorption
but little, or no, optical absorption) are the
Broad Absorption Line QSOs and the grism
selected QSOs.

The samples used are probably affected by selection
effects, which are discussed in the body of the paper, and therefore cannot
be considered as representative of the whole population of AGNs.
However, we can certainly state that at least
a sub-population of the AGNs is
characterized by a low $\rm E_{B-V}/N_H$ or $\rm A_V/N_H$ with respect to
the Galactic value. Also, we could {\it not} find evidence for a Galactic
standard $\rm E_{B-V}/N_H$ or $\rm A_V/N_H$ ratio in nearly any object.

We presented additional evidences indicating that the properties of dust grains
in the circumnuclear region of AGNs are different with respect to the Galactic
diffuse interstellar medium:\\
-- although type 2 Seyfert nuclei appear significantly absorbed by dust in the
mid--IR, their average ISO spectrum does not show evidence for the
silicate absorption feature at 9.7$\mu$m which, instead, is expected to be very
deep in case of heavy absorption;\\
-- some type 1 Seyferts which appear affected
by some reddening do not show evidence for the carbon dip at 2175\AA ,
while according to the Galactic standard extinction curve this absorption
feature should be prominent in their UV spectra.\\
Both these observational evidences suggest that dust in the circumnuclear
region of AGNs is depleted of the small grains ($<3\mu$m and 100-200\AA \
respectively) which are responsible for these absorption features.
A dust grain distribution biased in favor of large grains would also make the
extinction curve flatter (Laor \& Draine 1993). If the bias for large grains is
due to coagulation this would also explain the reduced $\rm E_{B-V}/N_H$ and
$\rm A_V/N_H$ (Kim \& Martin 1996). 
A more detailed discussion on the interpretation of these results is given in a
companion paper (Maiolino et al. 2000b, paper II).

Regardless of the intepretation of these observational phenomena, these results
have important consequences on the unified theories of AGNs. The finding that
the dust absorption in generally significanlty lower with respect to what
expected from the gaseous column density implies that for several AGNs
the optical classification
might be de-coupled from the X-ray classification. In particular, AGNs which
appear obscured (type 2) in the X-rays might appear as relatively unobscured
(type 1) in the optical.

\begin{acknowledgements}
We are grateful to B. Draine for enlightening discussions during the early
stages of this work. This paper also benefits of useful comments from
A. Natta, M. Walmsley and from the referee R. Antonucci.
This work was partially supported by the Italian Space Agency
(ASI) under grant ARS--99--15 and by the Italian Ministry for
University and Research (MURST) under grant Cofin98--02--32.
\end{acknowledgements}


\begin{thebibliography}{}

\bibitem[1994]{aguero} Aguero E L., Calderon J. H., Paolantonio S., Suarez Boedo E.,
	1994, PASP 106, 978

\bibitem[2000]{akiyama} Akiyama M., Ohta K., Yamada T., 2000, ApJ
	532, 700

\bibitem[1993]{anto} Antonucci R., 1993, ARA\&A 31, 473

\bibitem[1991]{awaki} Awaki H., Koyama K., Inoue H., Halpern J.P.,
	1991, PASJ 32, 195

\bibitem[1999]{bassani} Bassani L., Dadina M., Maiolino R., et al.,
	1999, ApJS 121, 473

\bibitem[1998]{boyle} Boyle B. J., Almaini O., Georgantopoulos I., et al., 
	1998, MNRAS 297, L53

\bibitem[1978]{bohlin} Bohlin R.C., Savage B.D., Drake J.F., 1978,
	ApJ 224, 132

\bibitem[1996]{brandt} Brandt W.N., Fabian A.C., Pounds K.A., 1996,
	MNRAS 278, 326

\bibitem[2000]{brandt} Brandt W.N., Laor A., Wills B.J., 2000a, ApJ 528,
	637

\bibitem[2000]{brandt} Brandt W.N., et al., 2000b, AJ, 119, 2349




\bibitem[2000]{clavel} Clavel J., Schulz B., Altieri B., et al., 2000,
 A\&A 357, 839


\bibitem[2000]{comastri} Comastri A., et al., 2000, proceedings of the conference
        ``X-ray astronomy 1999'', Astroph. Lett. and Comm., in press

\bibitem[1988]{crenshaw} Crenshaw D. M., Peterson B. M., Wagner R. M.,
	1988, AJ 96, 1208
	



\bibitem[1999]{feldmeier} Feldmeier J.J., Brandt W.N., Elvis M., et al., 1999,
	ApJ 510, 167


\bibitem[1988]{filippenko} Filippenko A., Sargent W.L.W., 1988, ApJ 324, 134

\bibitem[1999]{fiore} Fiore F., La Franca F., Giommi P., 
	et al., 1999, MNRAS 306, L55

\bibitem[2000]{fiore} Fiore F., et al., 2000, New Astronomy, in press


\bibitem[1991]{francis} Francis P.J., Hewett P. C., Foltz C. B., 
	et al., 1991, ApJ 373, 465

\bibitem[1999]{gall} Gallagher S.C., Brandt W.N., Sambruna R.M.,
	Mathur S., Yamasaki N., 1999, ApJ 519, 549

\bibitem[1998]{george} George I.M., Turner T.J., Netzer H., et al., 1998,
	ApJS 114, 73


\bibitem[2000]{gilli} Gilli R., Maiolino R., Marconi A., et al., 2000,
	A\&A 355, 485

\bibitem[1995]{goodrich} Goodrich R.W., 1995, ApJ 440, 141

\bibitem[1997]{granato} Granato G.L., Danese L., Franceschini A.,
 1997, ApJ 486, 147

\bibitem[1999]{halpern} Halpern J.P., Turner T.J., George I.M., 
	MNRAS 307, L47



\bibitem[1999]{ho} Ho L.C., 1999, Adv. Sp. Res. 23, 813

\bibitem[1999]{ho2} Ho L.C., Ptak A., Terashima Y., 1999, ApJ 525, 168




\bibitem[1997]{komossa1} Komossa S., Fink H., 1997a, A\&A 322, 719

\bibitem[1997]{komossa2} Komossa S., Fink H., 1997b, A\&A 327, 483

\bibitem[1998]{komossa3} Komossa S., Bade N., 1998, A\&A 330, 823

\bibitem[1995]{koratkar} Koratkar A., Deustua S. E., Heckman T., et al.,
	1995, ApJ 440, 132

\bibitem[1999]{kraemer} Kraemer S.B., Ho L.C., Crenshaw D.M., Shields J.C.,
	Filippenko A.V., 1999, ApJ 520, 564

\bibitem[2000]{kraemer} Kraemer S.B., George I.M., Turner T.J., 
	Crenshaw D.M., 2000, ApJ 535, 53

\bibitem[1982]{lacy} Lacy J.H., Soifer B.T., Neugebauer G., et al., 1982,
  ApJ 256, 75


\bibitem[1997]{leighly} Leighly K.M., Kay L.E., Wills B.J., Wills D., Grupe D.,
	1997, ApJ 489, L137

\bibitem[1982]{macca} Maccacaro T., Perola G.C., Elvis M., 1982,
	ApJ 257, 47

\bibitem[1994]{maiolino} Maiolino R., Rieke G.H., Rieke M.J., 1996,
  AJ, 111, 573

\bibitem[1995]{maiolino} Maiolino R., Rieke G.H., 1995, ApJ 454, 95

\bibitem[1998]{maiolino} Maiolino R., Salvati M., Bassani L., et al., 1998,
        A\&A 338, 781

\bibitem[2000]{maiolino} Maiolino R., 2000, in
	``X-ray astronomy '999'', Astroph. Lett. and Comm., in press
	(astro-ph/0007473)

\bibitem[2000]{maiolino} Maiolino R., Salvati M., Antonelli L.A., et al., 2000a,
  A\&A 355, L47

\bibitem[2000]{maiolino} Maiolino R., Marconi A., Oliva E.,
	2000b, A\&A, submitted (paper II)


	217, 425

\bibitem[1986]{mcalary} McAlary C.W., Rieke G.H., Lebofsky M.J.,
	Stocke J.T., 1986, ApJ 301, 105




\bibitem[1984]{mushotzky} Mushotzky R.F., Ferland G.J., 1984, ApJ 278, 558

\bibitem[2000]{mushotzky} Mushotzky R.F., Cowie L.L., Barger A.J., Arnaud K.A.,
	2000, Nature 404, 459

\bibitem[2000]{netzer1} Netzer H., 2000,
in ``X-ray astronomy 1999'', Astroph. Lett. and Comm., in press

\bibitem[1985]{netzer2} Netzer H., Helitzur M., Ferland G. J., 1985, ApJ 299, 752




\bibitem[1993]{pier} Pier E.A., Krolik J.H., 1993, ApJ 418, 673

\bibitem[1981]{puetter} Puetter R.C., Smith H.E., Willner S.P.,
	Pipher J.L., 1981, ApJ 243, 345

\bibitem[1988]{rees} Rees M. J., Netzer H., Ferland G.J., 1989, ApJ, 347, 640
	Absorption Lines. Cambridge University Press



\bibitem[1995]{reynolds} Reynolds C.S., Fabian A.C., 1995, MNRAS 273, 1167

\bibitem[1997]{reynolds} Reynolds C.S., 1997, MNRAS 286, 513

\bibitem[2000]{risaliti} Risaliti G., et al., 2000, A\&A, submitt.

\bibitem[1999]{risaliti2} Risaliti G., Maiolino R., Salvati M., 1999,
	ApJ, 522, 157

\bibitem[1990]{rix} Rix H.-W., Rieke G., Rieke M., Carleton N.P., 1990, ApJ 363, 480




\bibitem[1999]{sambruna} Sambruna R.M., Eracleous M., Mushotzky R.F.,
	1999, ApJ 526, 60

\bibitem[1979]{savage} Savage B. D., Mathis J. S., 1979, ARA\&A 17, 73

\bibitem[1989]{setti} Setti G., Woltjer L., 1989, A\&A 224, L21

\bibitem[1990]{stirpe} Stirpe G. M., 1990, A\&AS 85, 1049

\bibitem[1997]{turner} Turner T.J., George I.M., Nandra K.,
	Mushotzky R.F., 1997, ApJS 113, 23

\bibitem[1999]{turner} Turner T.J., 1999, ApJ, 511, 142


\bibitem[1997]{veilleux} Veilleux S., Goodrich R.W., Hill G.J., 1997, ApJ 477, 631

\bibitem[2000]{vignali} Vignali C., Mignoli M., Comastri A., Maiolino R., Fiore F.,
	2000, MNRAS 314, L11




\bibitem[2000]{weingartner} Weingertner, J.C., Draine, B.T., 2000, ApJ,
	in press (astro-ph/0008146)

\bibitem[1992]{winkler} Winkler H., Glass I. S., van Wyk F., 
	et al., 1992, MNRAS 257, 677


\end{thebibliography}
\end{document}